\documentclass[aps,a4paper,showpacs,nofootinbib]{revtex4}
\usepackage{graphicx}
\usepackage{color}
\usepackage{amsmath}
\usepackage{amssymb}
\usepackage{enumerate}
\usepackage{graphics,epsfig,subfigure}
\usepackage{epstopdf}
\usepackage{color}

\newcommand\be{\begin{equation}}
\newcommand\ee{\end{equation}}
\newcommand\ba{\begin{eqnarray}}
\newcommand\ea{\end{eqnarray}}
\newcommand\nn{\nonumber}

\begin{document}

\title{Stable Palatini $f(\mathcal{R})$ braneworld}
\author{Bao-Min Gu$^{1,2}$\footnote{baomin.gu@mail.mcgill.ca}, Yu-Xiao Liu$^{1}$ \footnote{liuyx@lzu.edu.cn, corresponding author}, Yuan Zhong$^{3}$\footnote{zhongy@mail.xjtu.edu.cn} }

\affiliation{$^1$Institute of Theoretical Physics \& Research Center of Gravitation
\& Key Laboratory for Magnetism and Magnetic Materials of the Ministry of Education, Lanzhou University, Lanzhou 730000, China}
\affiliation{$^2$ Department of Physics, McGill University, Montr\'{e}al, QC, H3A 2T8, Canada}
\affiliation{$^3$ School of Science, Xi'an Jiaotong University, Xi'an 710049, China}

\begin{abstract}
We consider the static domain wall braneworld scenario constructed from the Palatini formalism $f(\mathcal{R})$ theory. We check the self-consistency under scalar perturbations. By using the scalar-tensor formalism we avoid dealing with the higher-order equations. We develop the techniques to deal with the coupled system. We show that under some conditions, the scalar perturbation simply oscillates with time, which guarantees the stability. We also discuss the localization condition of the scalar mode by analyzing the effective
potential and the fifth dimensional profile of the scalar mode. We apply these results to an explicit example, and show that only some of the solutions allow for stable scalar perturbations. These stable solutions also give nonlocalizable massless mode. This is important for reproducing a viable four-dimensional gravity.
\end{abstract}

\pacs{04.50.Kd, 98.80.-k}
\maketitle



\section{Introduction and motivations}
The idea that extra spatial dimensions may exist has opened up a new gate towards new physics beyond the standard model of particle physics and of cosmology. It provides the possibility for solving some open problems such as the hierarchy problem \cite{Arkani-Hamed1998a,Randall1999}, the neutrino mass problem \cite{ArkaniHamed1998vp},  the fermion mass hierarchy problem \cite{Gherghetta:2000qt}, etc. It also provides some new explanation for  dark matter \cite{Cheng:2002ej} and dark energy \cite{Sahni:2002dx}. These features make the extra dimension theories attractive.

As is well known, gravity propagates differently in higher dimensions. In the large extra dimension theory \cite{Arkani-Hamed1998a}, gravity violates the Newton's inverse-square law at short distance below the extra dimension radius and recovers its four-dimensional feature at large distance. Hence experiments for testing the Newton's law at short distance are important for probing the extra dimensions \cite{Hoyle:2000cv,Adelberger:2003zx,Tan2016vwu}. In the Randall-Sundrum model \cite{Randall1999}, however, the situation is completely different: it is the massless Kaluza-Klein mode of graviton that accounts for the inverse-square law. It was found that the four-dimensional gravity can be recovered when the massless Kaluza--Klein mode is localized near the hidden brane. However, in the original proposal of Randall--Sundrum model, the size of the fifth dimension is not dynamically determined. In other words, the perturbation of the radius, the radion, is not stabilized. So the size of extra dimension which accounts for the hierarchy problem is fixed artificially, which is not natural and not stable. The absence of stabilization also makes the radion  massless. This is evidently unacceptable since it leads to a long-range fifth force which has never been observed. These problems can be solved by the Goldberger--Wise mechanism \cite{Goldberger:1999uk}. The radion is stabilized by introducing a bulk scalar field, whose vacuum expectation value is related to the extra dimension coordinate. Thus the radius can be dynamically fixed. Once this stabilization mechanism is introduced, the radion then becomes massive if backreaction is considered \cite{Csaki:2000zn}, and the modular can be dynamically fixed such that the hierarchy problem is naturally addressed.

In the smooth version of warped braneworld models \cite{DeWolfe2000,Csaki2000a,Kobayashi2001jd,Giovannini2001a,Giovannini2001b,
Bazeia:2003aw,Liu:2009ega}, the warping along the fifth dimension is caused by background scalar field, and the extra dimension is infinitely large. So there is no need for stabilization. But the infinite range of the extra dimension also implies the existence of massless  radion-like mode, in spite of the existence of bulk scalar field. The behavior of this massless mode is background-dependent. If the massless mode is localized, then it couples to the trace of the energy-momentum tensor of standard model particles. This is unviable since it leads to a long-range interaction. It is really a problem for domain wall braneworld models.
This is one of our motivation of this work. We expect to study the scalar mode in the domain wall braneworld model in Palatini $f(\mathcal{R})$ theory, and get the constraints under which the theory is phenomenologically consistent.

In addition to the localization problem, there is another problem. As a kind of toy model that allows for infinitely large extra dimension, the domain wall braneworld models are usually assumed to be static and have four-dimensional Poincar\'{e} symmetry, so that the background fields and the spacetime metric depend only on the fifth dimension. However, such kind of static model may not be consistent with time evolution. The model would be unstable if the perturbation grows with time. This is the stability problem of static braneworld models. It is usually referred to the tachyon instability problem since the instability can be described by the four-dimensional mass of the Kaluza--Klein modes. Clearly, physically viable models should be free of tachyon. Thus it is necessary to investigate the stability of the perturbations.

The Palatini $f(\mathcal{R})$ domain wall braneworld model has been considered in previous literature \cite{Bazeia:2014poa,Gu2014}. The exact domain wall solutions were obtained, and it was shown that the tensor perturbations are stable \cite{Gu2014}. The scalar perturbation remains unclear mostly because of the special structure of  Palatini $f(\mathcal{R})$ theory. The theory is assumed to have two independent variables, the spacetime metric and the independent connection. The connection can be eliminated thus one gets a metric theory, but with a modified source part. The complexity of the scalar perturbation mainly comes from this modified source. Such a theory has some special features on cosmology \cite{Meng:2003en,Flanagan2004,Olmo2005b,Koivisto2006,Amarzguioui2006,Fay:2007gg,
Koivisto2007a,Barausse2007,Barragan:2009sq,Barragan2010a,Olmo2011b}.
In braneworld scenario, it also has some interesting properties. In the previous braneworld models considered in other gravity theories \cite{Gremm2000a,Gremm2000,Adam:2007ag,Fu2014a,Bazeia2015}, the warp factor decays exponentially at the boundaries of the fifth dimension. This is because the warp factor is related to the localization condition of the massless graviton. Usually, growing warp factor solutions are not allowed since this would give nonlocalizable massless graviton and localizable massless scalar mode. However, the warp factor in Palatini $f(\mathcal{R})$ theory allows for both of decaying and growing solutions, and they all give localizable massless graviton \cite{Gu2014}. This may provide some new mechanisms to localize standard model particle fields. In this work, we will deal with the scalar perturbations by using scalar-tensor theory since it is widely accepted that $f(R)$ theory (both of metric formalism and Palatini formalism) has a mathematical equivalence with scalar-tensor theory.

The paper is organized as follows. In section \ref{section modelsetup} we give a model setup of the Palatini $f(\mathcal{R})$ theory, and show how to remove the connection dependence. In section \ref{section perturbation} we give the scalar-tensor formalism of the Palatini $f(\mathcal{R})$ theory, and develop the techniques to deal with the perturbations of nonminimally coupled theory. The equations of perturbations are obtained for both of the single field theory and two-field theory. In section \ref{section localization and stability} we analyze the localization problem and the stability against time evolution. At last, we give the conclusions in section \ref{section conclusion}.

\section{Model setup}\label{section modelsetup}
We start from the general Palatini formalism $f(\mathcal{R})$ theory with a background scalar field $\chi$,
\begin{equation}
S=\frac{1}{2\kappa_D}\int\mathrm{d}^Dx \sqrt{-g}f(\mathcal{R}(g,\Gamma))
+\int\mathrm{d}^Dx\sqrt{-g}\left(-\frac{1}{2}\partial_{M}\chi\partial^{M}\chi
-V(\chi)\right),
\label{action}
\end{equation}
where  $\mathcal{R}=g^{MN}\mathcal{R}_{MN}(\Gamma)$, and the Ricci tensor $\mathcal{R}_{MN}(\Gamma)$ is defined by
\begin{equation}
\mathcal{R}_{MN}(\Gamma)\equiv \partial_P \Gamma^{P}_{MN}
-\partial_N \Gamma^{P}_{MP}
+\Gamma^{P}_{PQ}\Gamma^{Q}_{MN}
-\Gamma^{P}_{MQ}\Gamma^{Q}_{P N}.
\end{equation}
The connection $\Gamma$ is not the Christoffel symbol constructed from the spacetime metric, but an independent variable. Note that the source field only couples to the spacetime metric.  One can immediately get the field equations for the metric,
\begin{equation}
f_{\mathcal{R}}\mathcal{R}_{MN}-\frac{1}{2}f(\mathcal{R})g_{MN}=\kappa_D T_{MN},
\label{original equation}
\end{equation}
and for the connection,
\begin{equation}
\widetilde{\nabla}_P\left(\sqrt{-g}f_{\mathcal{R}}g^{MN}\right)=0.
\label{gradient}
\end{equation}
The covariant derivative $\widetilde{\nabla}$ is compatible with the connection $\Gamma$. This formula has an analogy to that of general relativity, in which $\nabla_P g^{MN}=0$ (and $\nabla_P\left(\sqrt{-g}g^{MN}\right)=0$).
Indeed, the condition (\ref{gradient}) allows one to define an auxiliary metric $q_{MN}$ such that
\begin{equation}
\widetilde{\nabla}_P(\sqrt{-q}q^{MN})=0.
\end{equation}
The comparison with (\ref{gradient}) gives the solution $q_{MN}=f_{\mathcal{R}}^{\frac{2}{D-2}}g_{MN}$, which is just a conformal transformation of the spacetime metric. Now it can be easily checked that this auxiliary metric is the one that defines the connection $\Gamma$ and the covariant derivative $\widetilde{\nabla}$. Once again, as we mentioned in previous context, the source field couples to $g_{MN}$ other than $q_{MN}$. This is somewhat an assumption, however, which implies that the independent connection $\Gamma(q_{MN})$ does not define the spacetime parallel transport since the covariant derivatives in the source part are defined by the Christoffel symbol. In this sense, Palatini $f(\mathcal{R})$ theory is also a metric theory \cite{Sotiriou2010}. To be specific, expressing the Ricci scalar $\mathcal{R}$ in terms of $q_{MN}$ one gets the transformation
\begin{equation}
\mathcal{R}=R-\frac{2(D-1)}{(D-2)f_{\mathcal{R}}}
\nabla_M\nabla^M f_{\mathcal{R}}+
\frac{D-1}{(D-2)f_{\mathcal{R}}^2}\nabla_Mf_{\mathcal{R}}\nabla^M f_{\mathcal{R}},
\end{equation}
where $R$ is the usual curvature scalar defined by the spacetime metric. Thus the field equations (\ref{original equation}) and (\ref{gradient}) can be combined to get
\begin{eqnarray}
 G_{M N}&=&\frac{\kappa_{D} T_{M N}}{f_{\mathcal{R}}}-
 \frac{1}{2}g_{M N}\left(\mathcal{R}-\frac{f}{f_{\mathcal{R}}}\right)+
 \frac{1}{f_{\mathcal{R}}}\left(\nabla_{M}\nabla_{N}-g_{M N}\nabla_{A}\nabla^{A}\right)
 f_{\mathcal{R}}\nn\\
 &&-\frac{D-1}{(D-2)f_{\mathcal{R}}^{2}}
 \left(\nabla_{M}f_{\mathcal{R}}\nabla_{N}f_{\mathcal{R}}
 -\frac{1}{2}g_{M N}\nabla_{A}f_{\mathcal{R}}\nabla^{A}f_{\mathcal{R}}\right),
 \label{modified Einstein equation}
\end{eqnarray}
It seems that the  field equation still depends on the connection. However, there is a subtlety here. The trace of the equation (\ref{original equation}) gives an algebraic relation between $\mathcal{R}$ and the trace of the energy-momentum tensor. This implies that $\mathcal{R}$ is fully determined once we know the energy-momentum tensor (the source). Thus we see that the right-hand side of (\ref{modified Einstein equation}) is nothing but a modified source. This is actually a general feature of Palatini theories. The well-known Eddington-inspired Born-Infeld theory  \cite{Banados2010,Pani2012,Fu2014a,BeltranJimenez:2017doy} has a similar structure, but the formalism is much more complicated after the connection dependence is removed.

Since the energy-momentum tensor contains first-order derivatives, the modified source then contains third-order derivatives through $\mathcal{R}$, $f_\mathcal{R}$, and $f(\mathcal{R})$. For example, for scalar field we have $T\sim (\partial\phi)^2$, thus $\nabla^2 f_{\mathcal{R}}$ gives $\partial^3\phi\partial\phi$. Note that the term like $(\partial\phi)^2 \partial^2 g$ also appears. From this observation we see that the theory may give some special physics.

\section{Scalar perturbations in thick braneworld}\label{section perturbation}
\subsection{Scalar-tensor formalism}
Now let us turn to the braneworld perturbations of this theory. The tensor perturbations have been considered in Ref.~\cite{Gu2014}. The tensor modes satisfy a second-order equation, with a tiny modification from that of general relativity. This can be expected from Eq.~(\ref{modified Einstein equation}), in which all of the covariant derivatives on the right-hand side act on scalars. For scalar modes, as we have mentioned in previous section, there are third-order derivatives.

We start from Eq.~(\ref{modified Einstein equation}),  which can be derived from the action
\begin{equation}
S=\frac{1}{2\kappa_D}\int\mathrm{d}^Dx\sqrt{-g}\left(f_{\mathcal{R}}R
+\frac{D-1}{(D-2)f_{\mathcal{R}}}\partial_M f_{\mathcal{R}}\partial^M f_{\mathcal{R}}-\left(\mathcal{R}f_{\mathcal{R}}-f(\mathcal{R})\right)\right)
+\int\mathrm{d}^Dx\sqrt{-g}\mathcal{L}(g,\chi).
\label{Palatini action}
\end{equation}
Note that we dropped the total derivative terms.
Recall that there is an equivalence between $f(R)$ theory (both of the metric formalism and the Palatini formalism) and the scalar-tensor theory, which cast the higher-order theory into an ordinary second-order theory by introducing an extra scalar field. This scalar field actually describes the extra degree of freedom in theory with higher-order derivatives. For our model (\ref{Palatini action}), it is still a theory with higher-order derivatives if we regard the $\mathcal{R}$ (thus $f(\mathcal{R})$ and $f_{\mathcal{R}}$) terms as functions of the metric and the source $T$. Now we define $\phi\equiv f_{\mathcal{R}}$, then we have
\begin{equation}
S=\frac{1}{2\kappa_D}\int\mathrm{d}^Dx\sqrt{-g}\left(\phi R
+\frac{D-1}{(D-2)\phi}\partial_M \phi\partial^M \phi-V(\phi)\right)
+\int\mathrm{d}^Dx\sqrt{-g}\mathcal{L}(g,\chi),
\label{ST action}
\end{equation}
where $V(\phi)=\phi \mathcal{R}(\phi)-f(\mathcal{R}(\phi))$.  This is a theory without higher-order derivatives, but with one more degree of freedom. The field equation is then the equation (\ref{modified Einstein equation}) with $f_{\mathcal{R}}$ replaced by $\phi$.
The exact background solutions and the tensor perturbations were discussed in \cite{Gu2014}. It was shown that the tensor modes are stable under time evolution. The profile along the fifth dimension depends on the background, and there exists a localizable massless graviton.  However, the scalar modes remain unclear. We mainly deal with this problem in this work.

\subsection{Scalar perturbations}

Since we are considering the thick braneworld model, we require the five-dimensional background fields to have only $y$ dependence so that the four-dimensional Poincar\'{e} symmetry is conserved, i.e. $\phi_0\equiv \phi_0(y)$, $\chi_0 \equiv \chi_0(y)$.
For the scalar perturbations, we will work in the longitudinal gauge,
\begin{equation}
\mathrm{d}s^2=e^{2A(y)}\left[1+\Phi(x^{\sigma},y)\right]
\eta_{\mu\nu}\mathrm{d}x^{\mu}\mathrm{d}x^{\nu}
+\left[1+\Psi(x^{\sigma},y)\right]\mathrm{d}y^2.
\end{equation}
The perturbations of the scalar fields are defined as
\begin{equation}
\delta\phi(x^{\mu},y)\equiv\phi-\phi_0 (y),\quad
\delta\chi(x^{\mu},y)\equiv\chi-\chi_0 (y),
\end{equation}
where the lower index $0$ implies the background quantities.
The $5\mu$ part and the off-diagonal part of the $\mu\nu$ components of the perturbed equation (\ref{modified Einstein equation}) simply give two constraints on the scalar modes,
\begin{eqnarray}
\Psi&=&-\frac{\delta\phi}{\phi_0}-2\Phi,
\label{constraint 1}
\\
2\kappa_5 \chi_0' \delta\chi&=&2\left(3\phi_0 A' + \phi_0'\right)\Psi
+\left(2A'+\frac{8\phi_0'}{3\phi_0}\right)\delta\phi -2\delta\phi'-6\phi_0\Phi'   .\label{constraint 2}
\end{eqnarray}
Note that the constraint (\ref{constraint 1}) has the anisotropic contribution coming from the effective energy momentum tensor, which is absent in general relativity. The scalar modes $\Phi$ and $\delta\phi$ couple to each other, hence  one can only chose to eliminate the scalar modes  $\Psi$ and $\delta\chi$. However, this would lead to coupled perturbation equations. There is a novel technique that can largely simplify these constraints.
Let us consider the new variables
\begin{equation}
\widetilde{\Psi}=\Psi+a \frac{\delta\phi}{\phi_0}, \quad
\widetilde{\Phi}=\Phi+\frac{1-a}{2}\frac{\delta\phi}{\phi_0},
\label{technique}
\end{equation}
such that the constraint (\ref{constraint 1}) becomes
\begin{equation}
\widetilde{\Psi}+2\widetilde{\Phi}=0.\label{New constraint 1}
\end{equation}
Here $a$ is a dimensionless parameter. In terms of the new variables, the constraint (\ref{constraint 2}) can be expressed as
\begin{equation}
\frac{2\kappa_5 \chi_0'}{\phi_0} \delta\chi=
2\left(3A'+\frac{\phi_0 '}{\phi_0} \right)\widetilde{\Psi}-6\widetilde{\Phi}'
+\left(2(1-3a)A' - \left(\frac{1}{3}-a\right)\frac{\phi_0 '}{\phi_0}\right)
\frac{\delta\phi}{\phi_0}-(3a-1)\frac{\delta\phi'}{\phi_0}.\label{new constraint 2}
\end{equation}
Now we set $a=1/3$, then the constraint becomes
\begin{equation}
\frac{2\kappa_5 \chi_0'}{\phi_0} \delta\chi=
2\left(3A'+\frac{\phi_0 '}{\phi_0} \right)\widetilde{\Psi}-6\widetilde{\Phi}'.
\label{new constraint 3}
\end{equation}
Clearly, the constraints (\ref{New constraint 1}) and (\ref{new constraint 3}) have the same formalism as those obtained in general relativity.

\subsubsection{Single field}
We first consider the single field case, namely, $\mathcal{L}(g,\chi)=0$. The constraint (\ref{new constraint 3}) is simply
\begin{equation}
2\left(3A'+\frac{\phi_0 '}{\phi_0} \right)\widetilde{\Psi}-6\widetilde{\Phi}'
=0.
\end{equation}
It can be used to solve the mode $\widetilde{\Phi}$ and the solution is
\begin{equation}
\widetilde{\Phi}(x^\sigma,y)=e^{-2A} \phi_0^{-2/3}\epsilon(x^\sigma).
\end{equation}
An intuitive idea is to solve the other one scalar mode $\delta\phi$ with the perturbation equations. However, it can be easily checked that the perturbation equations reduce to be $0\times\delta\phi=m^2\phi_0(y)$.
This means that it is impossible to solve $\delta\phi$. Instead, we get a constraint, $m^2\phi_0(y)=0$ with $m^2=\Box\epsilon(x^\sigma)/{\epsilon(x^\sigma)}$. The solution is $m^2=0$ or $\phi_0(y)=0$. Using the constraint (\ref{technique}) we show that
the solution $\phi_0(y)=0$ leads to $\delta\phi=0$ and the divergence of $\Phi$, which is unviable, hence we have $m^2=0$ and $\phi_0(y)\neq0$. This implies that there exists only a massless Kaluza--Klein mode. By considering the background equations we show that the only solution is
\begin{equation}
A(y)\propto y, \quad \phi_0(y)=\text{constant}, \quad  V(\phi)=0.
\end{equation}
This is the model of general relativity with a cosmological constant, which is exactly the Randall-Sundrum model \cite{Randall1999a} if we insert a thin brane at the origin. The scalar perturbation $\widetilde{\Phi}$ is sort of a radion-like mode. There is no need to stabilize this mode since the extra dimension is infinitely large. One can easily show that this massless mode cannot be localized. So there is no extra long-range force contributing to the four-dimensional gravity.  Recall that the single field case corresponds to the Palatini $f(R)$ theory without source. Hence, we conclude that it is impossible to get a thick braneworld model in pure geometric Palatini $f(R)$ theory, and there are only thin brane solutions in this case.

\subsubsection{Two fields}

For general theory (\ref{Palatini action}), the scalar $\chi$ plays the role of source.
As can be seen from the constraints (\ref{New constraint 1}) and (\ref{new constraint 3}), we are not able to solve $\delta\phi$ in terms of $\widetilde{\Phi}$ and $\delta\chi$ in this case. So the only choice of variables to be eliminated is ($\widetilde{\Psi},\delta\chi$).
Varying the quadratic order of the action (\ref{ST action}) with respect to $\Psi$ and $\delta\phi$, and replacing
the scalar modes $\Phi$ and $\Psi$ with $\widetilde{\Phi}$ and $\widetilde{\Psi}$, we get the perturbation equations
\begin{eqnarray}
e^{2A}\widetilde{\Phi}''+\Box\widetilde{\Phi}+P_2(y) \widetilde{\Phi}'+Q_2(y)\widetilde{\Phi}=S_2(y)\delta\phi,\label{Palatini peq1}
\\
4e^{2A}\widetilde{\Phi}''+\Box\widetilde{\Phi}+P_3(y) \widetilde{\Phi}'+Q_3(y)\widetilde{\Phi}=S_3(y)\delta\phi.\label{Palatini peq2}
\end{eqnarray}
The coefficients $P_i$, $Q_i$, and $S_i$ are listed in the Appendix section. We  find that there is only one independent scalar mode in this theory. Clearly,  $\delta\phi$ has the same mass spectrum with $\widetilde{\Phi}$, so we only need to discuss $\widetilde{\Phi}$. It satisfies
\begin{equation}
e^{2A}\frac{S_3-4S_2}{S_3-S_2}\widetilde{\Phi}''
+\Box\widetilde{\Phi}+\frac{S_3 P_2-S_2 P_3}{S_3-S_2}\widetilde{\Phi}'
+\frac{S_3 Q_2-S_2 Q_3}{S_3-S_2}\widetilde{\Phi}=0.
\label{perturbation eq}
\end{equation}
It has a Shr\"{o}dinger-like formalism of equation in  coordinate $r$,
\begin{equation}
\left(-\partial_r^2+V_{\text{eff}}(r)\right)\eta=m^2\eta.
\label{shrodinger eq}
\end{equation}
The coordinate $r$ is defined by $dy=\zeta dr$. The variable $\eta$ is defined by the decomposition $\widetilde{\Phi}(x^\sigma,r)=\epsilon(x^\sigma)\gamma(r)\eta(r)$ in such a way that the first order derivative term can be eliminated. Note that we used the relation $\Box\epsilon(x^\sigma)=m^2\epsilon(x^\sigma)$.
There are two reasons why we write this formalism equation. First, we want to know how does the background affect the mass spectrum, and the Shr\"{o}dinger-like equation helps.
Second, this formalism implies that the quantity $\upsilon=\epsilon(x^\sigma)\eta(r)$ is the canonically normalized scalar mode. In other words, the quadratic order perturbation of action  (\ref{ST action}) is simply
\begin{equation}
S^{(2)}=\frac{1}{2}\int\mathrm{d}^4x\mathrm{d}r \left(\upsilon\partial_r^2\upsilon+
\upsilon\Box\upsilon-V_{\text{eff}}\upsilon^2\right).
\end{equation}
The expressions of $\gamma(r)$ and $\zeta(r)$ are
\begin{equation}
\zeta(r)=e^{A}\sqrt{\frac{S_3-4S_2}{S_3-S_2}},\quad
\gamma(r)=\zeta(r)\times\text{exp}\left(-\int \mathrm{d}r
\frac{S_3 P_2-S_2 P_3}{2e^{A}\sqrt{(S_3-S_2)(S_3-4S_2)}}\right).
\label{transformation}
\end{equation}
Here we have a constraint for the background solutions, $(S_3-S_2)(S_3-4S_2)>0$. If it is negative then the coordinate transformation from $y$ to $r$ does not hold. In fact, there would be a gradient instability if $\zeta(r)$ is imaginary, so this is not viable. Now we can write the effective potential in Eq.~(\ref{shrodinger eq}) as
\begin{equation}
V_{\text{eff}}(r)=2\left(\frac{\partial_r \gamma}{\gamma}\right)^2-\frac{\partial_r^2 \gamma}{\gamma}-\frac{S_3 Q_2-S_2 Q_3}{S_3-S_2}.
\label{shrodinger eq eff}
\end{equation}

\section{Stability problem and localization problem}
\label{section localization and stability}
We have mentioned the stability problem in the introduction section. Now let us explain this problem in detail.  Recall that we used the relation $\Box\epsilon(x^\sigma)=m^2\epsilon(x^\sigma)$ in Eq.~(\ref{shrodinger eq}). It was obtained by considering the fourier expansion in momentum space and using $p^\mu p_\mu=-m^2$. Clearly, $m$ plays the role of the four-dimensional mass of the scalar perturbation mode $\epsilon(x^\sigma)$, and the spectrum is determined by the background spacetime through (\ref{shrodinger eq eff}). This is actually an assumption of plane wave nature for the scalar mode $\epsilon(x^\sigma)$. However, it is really a problem whether it is consistent to assume that the scalar mode oscillates with time. The background system would be unstable if the scalar mode has nontrivial time evolution like growing or damping solutions. This instability can be described by the sign of $m^2$.  If there exist perturbation modes with $m^2<0$, then we would have an imaginary frequency $\omega=\sqrt{m^2+\vec{p}^2}$, which gives rise to instability and destroys the static background. Therefore, to get a consistent model it is necessary to have  nonnegative $m^2$. We see that this is actually a tachyon instability problem.

For tensor modes, it is straightforward to know that the effective potential supports a nonnegative $m^2$, since the perturbation equation can be factorized as $Q^\dag Q h_{\mu\nu}=m^2 h_{\mu\nu}$ \cite{Gu2014} for the theory with $f_{\mathcal{R}}>0$. In some other models considered in the previous literature \cite{Giovannini2001a,Kobayashi2001jd,Gu:2016nyo}, the effective potential for the scalar mode can also be factorized. However, for our case (\ref{shrodinger eq eff}), it is not clear whether it can be factorized for general Palatini $f(\mathcal{R})$ theory. If it can be factorized, then we have
\begin{equation}
V_\text{eff}(r)=\partial_r\left(\lambda-\frac{\partial_r \gamma}{\gamma}\right)
+\left(\lambda-\frac{\partial_r \gamma}{\gamma}\right)^2.
\end{equation}
Comparing this with Eq.~(\ref{shrodinger eq eff}), we get
\begin{equation}
\frac{S_3 Q_2-S_2 Q_3}{S_3-S_2}=2\lambda\frac{\partial_r \gamma}{\gamma}
-\partial_r\lambda-\lambda^2.
\label{condition 1}
\end{equation}
If this equation has regular solution for $\lambda(r)$, then we can factorize the Shr\"{o}dinger-like equation to be
\begin{equation}
\left(\partial_r+\lambda-\frac{\partial_r \gamma}{\gamma}\right)
\left(-\partial_r+\lambda-\frac{\partial_r \gamma}{\gamma}\right)\eta
=m^2\eta,
\label{factorized form}
\end{equation}
which obviously has the formalism $Q^\dag Q\eta=m^2\eta$ and implies $m^2\geq 0$. It also gives the solution for the massless scalar mode by
\begin{equation}
\left(-\partial_r+\lambda-\frac{\partial_r \gamma}{\gamma}\right)\eta_0=0.
\end{equation}
\begin{figure*}[htb]
\begin{center}
\includegraphics[width=8cm,height=6cm]{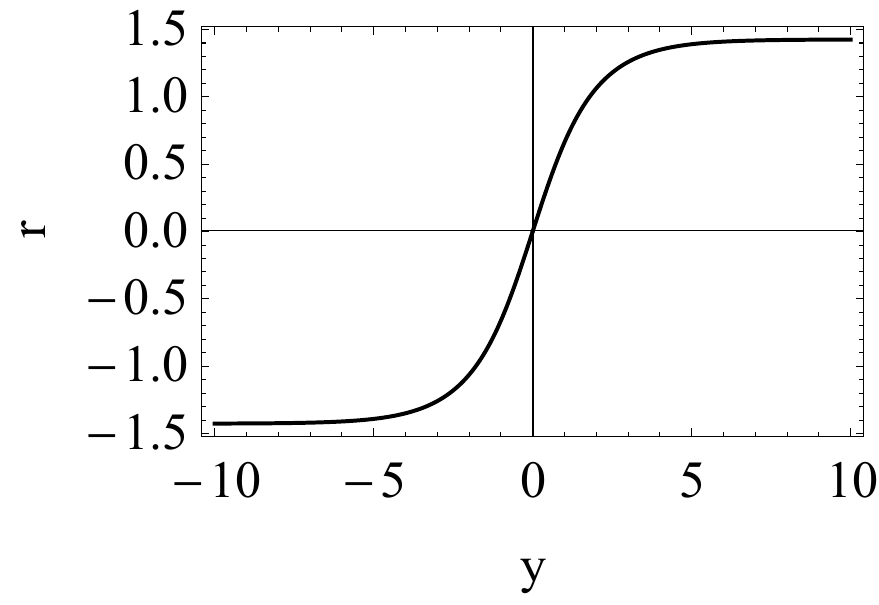}
\includegraphics[width=8.9cm,height=5.8cm]{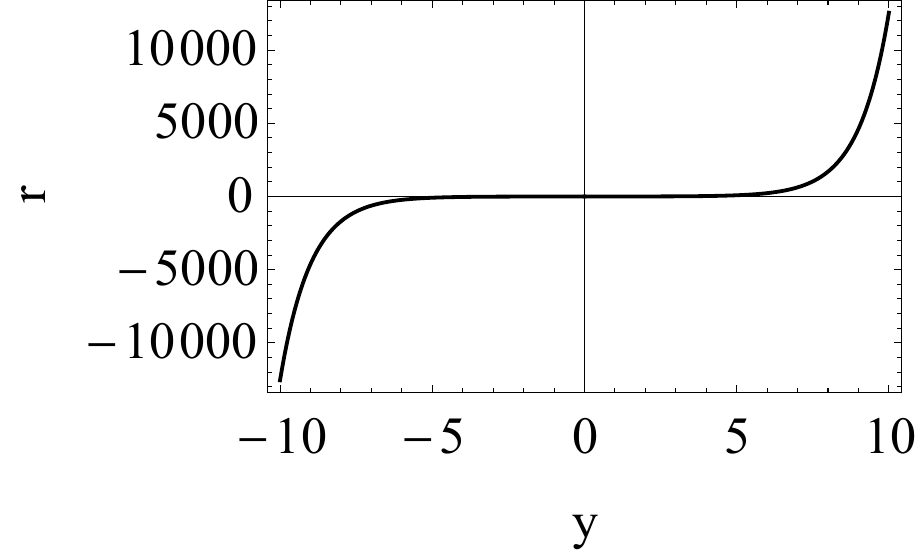}
\end{center}
\caption{The plot of the coordinate transformation $r(y)$ determined by the transformation function $\zeta$, with $n=1/7$ for the left and $n=5/3$ for the right. For $0<n<1/6$ the coordinate $r$ has a finite range, while for $n>1$ the coordinate $r$ has an infinite range. }\label{coord trans}
\end{figure*}
The localization condition of the massless scalar mode is
\begin{equation}
\int|\eta_0|^2\mathrm{d}r=\int|\eta_0/\sqrt{\zeta}|^2\mathrm{d}y<\infty.
\end{equation}
We see that it is the mode $\eta_0/\sqrt{\zeta}$ that describes the localization condition in $y$ coordinate.

There is another possibility which  also gives nonnegative $m^2$. Note that Eq.~(\ref{shrodinger eq eff}) can be written as
\begin{equation}
\left[\left(\partial_r-\frac{\partial_r \gamma}{\gamma}\right)
\left(-\partial_r-\frac{\partial_r \gamma}{\gamma}\right)
-\frac{S_3 Q_2-S_2 Q_3}{S_3-S_2}\right]\eta
=m^2\eta.
\label{special case}
\end{equation}
If
\begin{equation}
-\frac{S_3 Q_2-S_2 Q_3}{S_3-S_2}>0,
\label{condition 2}
\end{equation}
then the operator on the left-hand side gives a positive definite eigenvalue, i.e. $m^2>0$. This is also a possibility that avoids the tachyon instability problem. Note that  there is no massless scalar mode in this case.

Now let us apply the above discussion to the exact solution for Palatini $f(\mathcal{R})$ braneworld model given by reference \cite{Gu2014}. The model is $f(\mathcal{R})=\mathcal{R}+\alpha \mathcal{R}^2$, which is a simple modification to general relativity, and the modification is  described by the parameter $\alpha$. The solutions for the background quantities are
\begin{eqnarray}
 A(y)&=&\frac{2}{3(n-1)}\ln[\text{sech}(ky)],\\
 \phi(y)&=&\left(\frac{6n-1}{3n+2}\right)^{1/3} \text{sech}^{\frac{2n}{3(n-1)}}(ky),
 \\
 \chi(y)&=&\sqrt{\frac{2n(6n-1)}{3(3n+2)(n-1)\kappa_{5}}}
 \int\mathrm{d}y \sqrt{\text{sech}^4(ky)+2\text{sech}^2(ky)}.
 \label{background solutions}
\end{eqnarray}
To make the background solutions consistent, the parameter $n$ is restricted to be $n<-{2}/{3}$ or $0<n<{1}/{6}$ or $n>1$. With these solutions, we can compute the background quantities $P_i$, $Q_i$, and $S_i$ in Eq. (\ref{perturbation eq}). The explicit expression for the coordinate transformation factor in (\ref{transformation}) is
\begin{equation}
\zeta =\text{sech}^{\frac{2}{3(n-1)}}(ky) \sqrt{\frac{4(5 \cosh (2 k y)+3 n+7)}{(9 n+11) \cosh (2 k y)+30 n+10}}.
\end{equation}
This transformation gives the coordinate  $r=\int \mathrm{d}y\zeta(y)$. We give the plot of $r(y)$ in Fig. \ref{coord trans}. Note that if $\zeta$ is imaginary then we would have negative coefficient for $\widetilde{\Phi}$ and $\upsilon$, which leads to gradient instability.
To make the transformation regular, and to avoid gradient instability, we require $\zeta$ to be real. This requirement rules out the solution with $n<-2/3$. Further more, we have
\begin{equation}
-\frac{S_3 Q_2-S_2 Q_3}{S_3-S_2}=\frac{32 k^2 n (5 \cosh (2 k y)+3 n+7) \text{sech}^{\frac{4}{3 (n-1)}}(k y)}{(n-1) (\cosh (2 k y)+2) ((9 n+11)
\cosh (2 k y)+30 n+10)}.
\label{condition 2}
\end{equation}

\begin{figure*}[htb]
\begin{center}
\includegraphics[width=8.5cm,height=6.6cm]{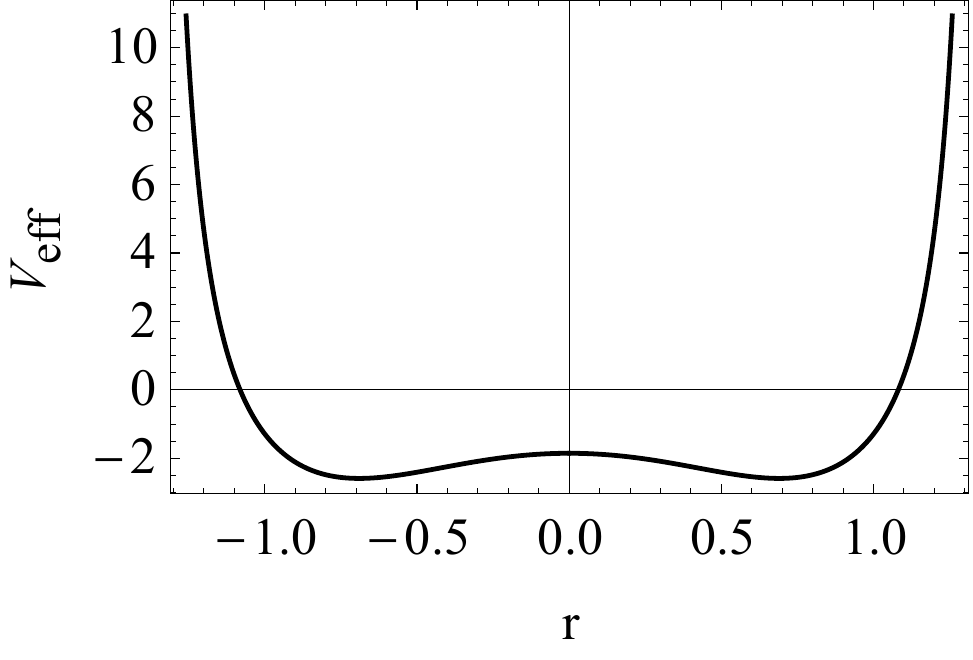}
\includegraphics[width=8.5cm,height=6.9cm]{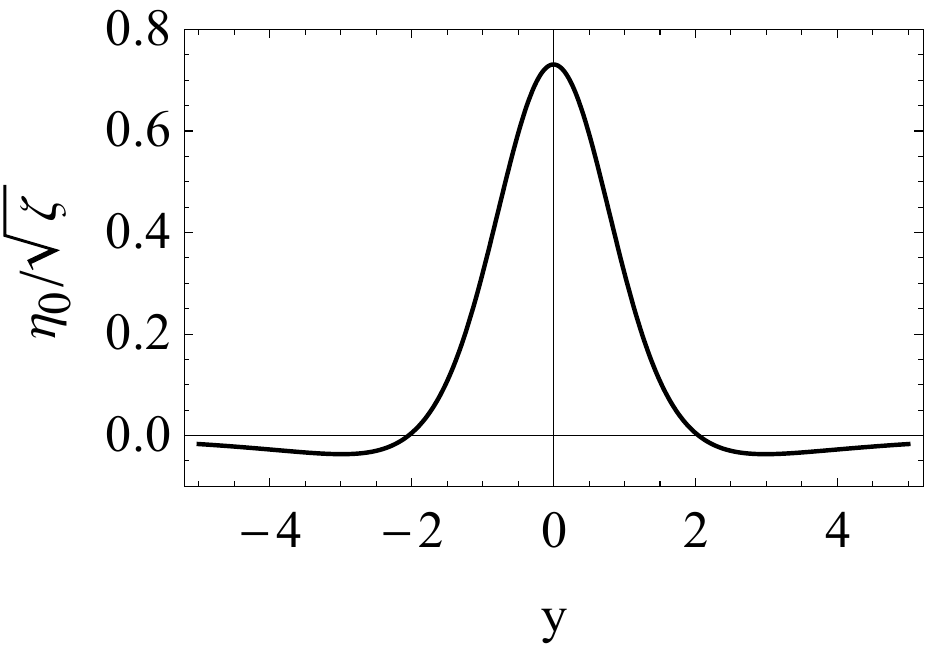}
\includegraphics[width=8.5cm,height=6.5cm]{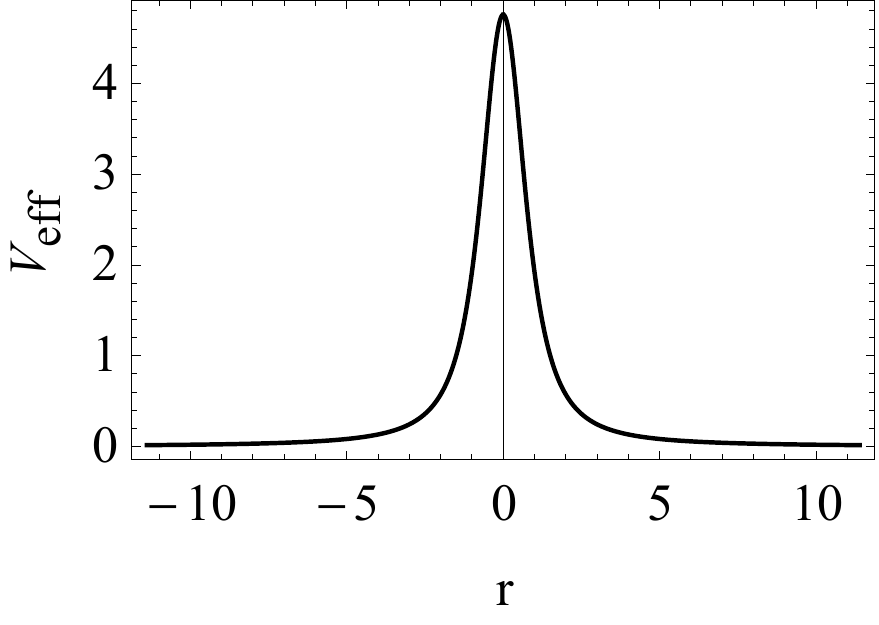}
\includegraphics[width=8.5cm,height=6.8cm]{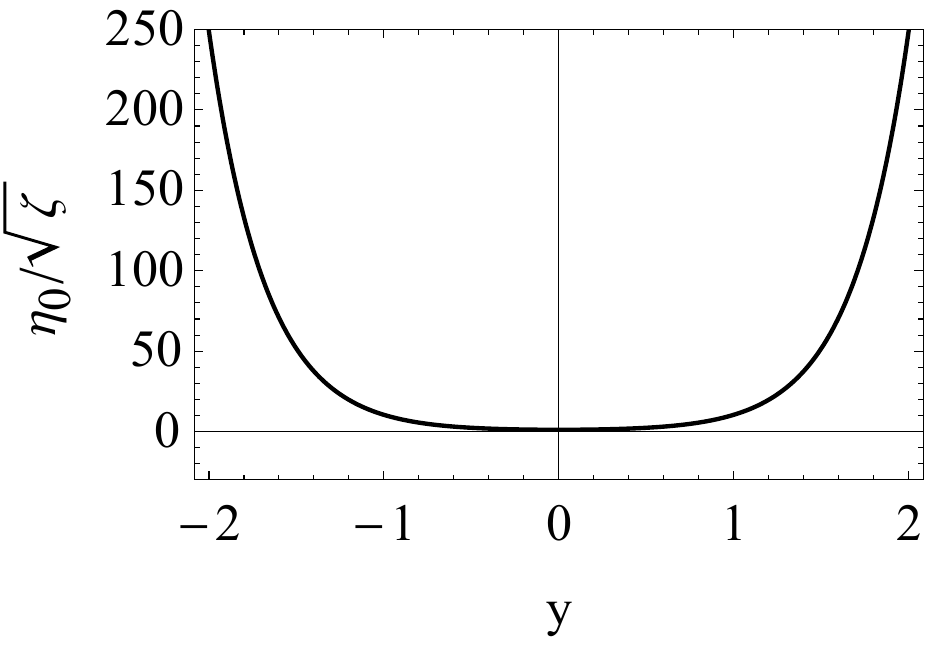}
\end{center}
\caption{The effective potential $V_{\text{eff}}(r)$ and the corresponding $\eta_0/\sqrt{\zeta}$ in $y$ coordinate, with $n=1/7$ (up) $n=5/3$ (bottom) respectively. For $n=1/7$, $V_{\text{eff}}(r)$ is an infinitely high potential well, and it has boundary since $r$ has finite range. The plot of $\eta_0/\sqrt{\zeta}$ reveals that the massless mode $\eta_0$ has nodes, so it is not the lowest state, which implies the existence of tachyon. For $n=5/3$, the potential is positive definite. The mode $\eta_0/\sqrt{\zeta}$ cannot be normalized.}\label{Vr and zero}
\end{figure*}

For $0<n<1/6$, the quantity (\ref{condition 2}) is negative and it may give negative $m^2$.  Let us check this by considering the effective potential $V_{\text{eff}}$ directly. The lengthy expression of $V_{\text{eff}}$ is given in the Appendix section. Note that we write it in $y$ coordinate, which does not affect our analyses. Clearly, for $0<n<1/6$, the potential blows up. We have $V_{\text{eff}}<0$ at the origin $y=0$ ($r=0$). In fact, since the transformation function defined in (\ref{transformation}) diverges, the new coordinate $r$ has a finite range.  It implies that $V_{\text{eff}}(r)$ is actually an infinitely  high potential well.  That means all of the scalar modes, including the massless mode, are localized.  This would contribute a long-range force to the gravity, and cause a violation to the observed four-dimensional Newtonian gravity, which is unacceptable. Hence the solution with  $0<n<1/6$ should also be ruled out. We give a typical plot of $V_{\text{eff}}(r)$ and the numerical solution of $\eta_0/\sqrt{\zeta}$ corresponding to the massless mode in Fig. \ref{Vr and zero}. Obviously, such a profile  is localizable. The nodes reveal the existence of lower states, the tachyon modes.  So the solutions with $0<n<1/6$ should also be ruled out. This result also implies that the effective potential may not be able to be factorized into the form (\ref{factorized form}), since if it can be factorized then there are no tachyon states.

For $n>1$, the effective potential is positive definite.  There is no scalar mode with $m^2<0$ in this case. Therefore, the model is stable under scalar perturbations. About the massless mode, we failed to get the analytic solution. But it is straightforward to conclude that it cannot be localized since the potential is positive definite. Hence there is no violation to the four-dimensional gravity. The plot of the potential and the numerical solution of $\eta_0/\sqrt{\zeta}$ corresponding to the massless mode are given in Fig. \ref{Vr and zero}. As can be seen, the massless mode diverges at infinity so it cannot be localized.

The above analysis shows that the model should be constrained if scalar perturbations are considered. This is one of our goal of this work. The previous work  is not enough to constrain the model. For the solutions given in \cite{Gu2014}, the constraints are $n>1$ and $\zeta^2>0$.

\section{conclusions}
\label{section conclusion}
To summarize, we studied the scalar perturbations of thick braneworld model in Palatini $f(\mathcal{R})$ theory. By  taking the advantages of scalar-tensor theory, we avoided dealing with the perturbations with third-order derivatives.
For a pure geometric theory, we showed that it is impossible to get a smooth version of braneworld model. This is contrary to the metric $f(R)$ theory \cite{Dzhunushaliev2010,Nojiri2011,Liu:2011am,Bazeia2014a,ZhongLiu2015}. For the theory with a source field, we used some techniques for the constraints, which can largely simplify the perturbation equations. In this case, there is only one independent scalar perturbation mode, although we have a nonminimally coupled scalar field $\phi$ and a source scalar field $\chi$.
Essentially, this is because the extra scalar degree of freedom is just an auxiliary field which is not dynamically independent.
This can also be understood from the fact that the
Palatini $f(\mathcal{R})$ theory corresponds to the Brans--Dicke theory with the Brans--Dicke parameter $\omega_{BD}=4/3$ in five dimensions, in which the nonminimally coupled scalar does not have its own dynamics \cite{Sotiriou2010}. So there is only one independent scalar degree of freedom in this system.

We also analyzed the stability problem and the localization problem. For a general theory, we failed to get a factorized formalism of perturbation equation. If it can be factorized into a formalism like $Q^\dag Q\eta=m^2\eta$, then we can conclude that there are no tachyon states, thus the scalar perturbation would be stable under time evolution, which makes the static system consistent. As an explicit example, we analyzed some  exact background solutions given in the previous literature. We showed that only the solutions with $n>1$ are stable and the other solutions should be ruled out if scalar perturbation is considered.  The corresponding massless mode cannot be localized, which guarantees the recovering of four-dimensional gravity. This gives some further constraints on the original Palatini $f(\mathcal{R})$ theory.

It is interesting to consider general scalar-tensor theory.
There are more degrees of freedom, so there are more independent  perturbation equations. Obviously, the scalar modes couple to each other. The braneworld models with  multiple scalar fields has been considered in reference \cite{Bazeia:2004dh,Aybat:2010sn,George:2011tn,Chen:2017diy}. It is interesting to consider the theory with nonminimally coupled scalar field, and find out the conditions under which the theory is viable, and that will be our future work.

\section{Acknowledgements}
We thank Robert Brandenberger for the helpful discussions.  This work was supported in part by the National Natural Science Foundation of China (Grants Nos. 11522541, 11375075, and 11605127), and the Fundamental Research Funds for the Central Universities (Grants Nos. lzujbky-2017-it69). Bao-Min Gu was supported by the scholarship granted by the Chinese Scholarship Council (CSC).

\appendix

\section{}

\subsection{The expressions of $P_i$, $Q_i$, $S_i$, and $V_{\text{eff}}(y)$}\label{appendix}
In the example of the two-field theory, the  coefficients in Eqs.~(\ref{Palatini peq1}) and (\ref{Palatini peq2}) are listed below:
\begin{eqnarray}
 P_2&=&e^{2A}\left(2A'+\frac{3\phi_0'}{\phi_0}-\frac{2\chi_0''}{\chi_0'}\right),
 \\
 Q_2&=&4e^{2A}\left(A'\frac{\phi_0'}{\phi_0}
 +A''+\frac{\phi_0''}{3\phi_0}
 -\left(A'+\frac{\phi_0'}{3\phi_0}\right)\frac{\chi_0''}{\chi_0'}\right),
 \\
 S_2&=&-\frac{e^{2A}}{9\phi_0}\left(4\frac{\phi_0'^2}{\phi_0^2}-9A''
 +3A'\frac{\phi_0'}{\phi_0}-3\frac{\phi_0''}{\phi_0}\right),
 \\
 P_3&=&4e^{2A}\left(7A'+2\frac{\phi_0'}{\phi_0}\right),
 \\
 Q_3&=&8e^{2A}\left(5A'^2+2A''-\frac{\phi_0'^2}{3\phi_0^2}+
 \frac{8}{3}\frac{\phi_0'}{\phi_0}A'+\frac{2}{3}\frac{\phi_0''}{\phi_0}\right),
 \\
  S_3&=&-4e^{2A}\left(5A'+2\frac{\chi_0''}{\chi_0'}\right)
 \left(\frac{A'}{3\phi_0}+\frac{4\phi_0'}{9\phi_0^2}
 -\frac{A''}{\phi_0'}-\frac{\phi_0''}{3\phi_0\phi_0'}\right).
\end{eqnarray}
The explicit expression for the effective potential is
\begin{eqnarray}
V_{\text{eff}}(y)&=&\bigg\{k^2 \big(2 \left(3 n \left(9 n \left(n \left(3888 n^2+70305 n+449950\right)+679790\right)+3147190\right)+1752239\right) \cosh (6 k y)
\nonumber \\
&+&25 \mathbf{(3 n-1)} (9 n+11)^2 \cosh (12 k y)-30 (n (3 n (9 n-293)-313)-235) (9 n+11) \cosh (10 k y)
\nonumber \\
&+&36 (n (9 n (n (n (38904 n+337921)+874186)+955526)+3275674)+245493) \cosh (2 k y)
\nonumber \\
&+&9 (3 n (n (3 n (128 n (249 n+2827)+1387381)+5291653)+2269913)+785321) \cosh (4 k y)
\nonumber \\
&+&30 (n (3 n (9 n (204 n+3895)+55289)+97939)+29521) \cosh (8 k y)+6 (3 n (3 n (n (12 n (21168 n+136177)+3540839)
\nonumber \\
&+&3664529)+4065547)+741787)\big) \text{sech}^{\frac{4}{3 (n-1)}}(k y)\bigg\}\bigg/
  \bigg\{24 \mathbf{(n-1)} \Big(\cosh (2 k y)+2\Big)^2 \Big(5 \cosh (2 k y)+3 n+7\Big)
\nonumber \\
&\times&\Big[(9 n+11) \cosh (2 k y)+30 n+10\Big]^3\bigg\}.
\end{eqnarray}
Note that the coefficient $3n-1$ of $\cosh (12 k y)$ in the numerator
 and the coefficient $n-1$ of the denominator determine the sign of the effective potential.


\begin{thebibliography}{10}

\bibitem{Arkani-Hamed1998a}
N.~Arkani-Hamed, S.~Dimopoulos, and G.~Dvali, {\it {The Hierarchy problem and
  new dimensions at a millimeter}},  {\em Phys.Lett.B} {\bf 429} (1998)
  263--272, [\href{http://arxiv.org/abs/hep-ph/9803315}{ hep-ph/9803315}].

\bibitem{Randall1999}
L.~Randall and R.~Sundrum, {\it {A Large mass hierarchy from a small extra
  dimension}},  {\em Phys.Rev.Lett.} {\bf 83} (1999) 3370--3373,
  [\href{http://arxiv.org/abs/hep-ph/9905221}{ hep-ph/9905221}].

\bibitem{ArkaniHamed1998vp}
N.~Arkani-Hamed, S.~Dimopoulos, G.~R. Dvali, and J.~March-Russell, {\it
  {Neutrino masses from large extra dimensions}},  {\em Phys.Rev.D} {\bf 65}
  (2001) 024032, [\href{http://arxiv.org/abs/hep-ph/9811448}{
  hep-ph/9811448}].

\bibitem{Gherghetta:2000qt}
T.~Gherghetta and A.~Pomarol, {\it {Bulk fields and supersymmetry in a slice of
  AdS}},  {\em Nucl.Phys.} {\bf B586} (2000) 141--162,
  [\href{http://arxiv.org/abs/hep-ph/0003129}{ hep-ph/0003129}].

\bibitem{Cheng:2002ej}
H.-C. Cheng, J.~L. Feng, and K.~T. Matchev, {\it {Kaluza-Klein dark matter}},
  {\em Phys.Rev.Lett.} {\bf 89} (2002) 211301,
  [\href{http://arxiv.org/abs/hep-ph/0207125}{ hep-ph/0207125}].

\bibitem{Sahni:2002dx}
V.~Sahni and Y.~Shtanov, {\it {Brane world models of dark energy}},  {\em JCAP}
  {\bf 0311} (2003) 014, [\href{http://arxiv.org/abs/astro-ph/0202346}{
  astro-ph/0202346}].

\bibitem{Hoyle:2000cv}
C.~D. Hoyle, U.~Schmidt, B.~R. Heckel, E.~G. Adelberger, J.~H. Gundlach, D.~J.
  Kapner, and H.~E. Swanson, {\it {Submillimeter tests of the gravitational
  inverse square law: a search for `large' extra dimensions}},  {\em Phys.Rev.Lett.} {\bf 86} (2001) 1418--1421,
  [\href{http://arxiv.org/abs/hep-ph/0011014}{ hep-ph/0011014}].

\bibitem{Adelberger:2003zx}
E.~G. Adelberger, B.~R. Heckel, and A.~E. Nelson, {\it {Tests of the
  gravitational inverse square law}},  {\em Ann.Rev.Nucl.Part.Sci.} {\bf
  53} (2003) 77--121, [\href{http://arxiv.org/abs/hep-ph/0307284}{
  hep-ph/0307284}].

\bibitem{Tan2016vwu}
W.-H. Tan, S.-Q. Yang, C.-G. Shao, J.~Li, A.-B. Du, B.-F. Zhan, Q.-L. Wang,
  P.-S. Luo, L.-C. Tu, and J.~Luo, {\it {New Test of the Gravitational
  Inverse-Square Law at the Submillimeter Range with Dual Modulation and
  Compensation}},  {\em Phys.Rev.Lett.} {\bf 116} (2016) 131101.

\bibitem{Goldberger:1999uk}
W.~D. Goldberger and M.~B. Wise, {\it {Modulus stabilization with bulk
  fields}},  {\em Phys.Rev.Lett.} {\bf 83} (1999) 4922--4925,
  [\href{http://arxiv.org/abs/hep-ph/9907447}{ hep-ph/9907447}].

\bibitem{Csaki:2000zn}
C.~Csaki, M.~L. Graesser, and G.~D. Kribs, {\it {Radion dynamics and
  electroweak physics}},  {\em Phys.Rev.D} {\bf63} (2001) 065002,
  [\href{http://arxiv.org/abs/hep-th/0008151}{ hep-th/0008151}].

\bibitem{DeWolfe2000}
O.~DeWolfe, D.~Z. Freedman, S.~S. Gubser, and A.~Karch, {\it Modeling the fifth dimension with scalars and gravity},  {\em Phys.Rev.D} {\bf 62} (2000)
  046008, [\href{http://arxiv.org/abs/hep-th/9909134}{ hep-th/9909134}].

\bibitem{Csaki2000a}
C.~Csaki, J.~Erlich, T.~J. Hollowood, and Y.~Shirman, {\it {Universal aspects
  of gravity localized on thick branes}},  {\em Nucl.Phys.B} {\bf 581} (2000)
  309--338, [\href{http://arxiv.org/abs/hep-th/0001033}{ hep-th/0001033}].

\bibitem{Kobayashi2001jd}
S.~Kobayashi, K.~Koyama, and J.~Soda, {\it {Thick brane worlds and their
  stability}},  {\em Phys.Rev.D} {\bf 65} (2002) 064014,
  [\href{http://arxiv.org/abs/hep-th/0107025}{ hep-th/0107025}].

\bibitem{Giovannini2001a}
M.~Giovannini, {\it Gauge-invariant fluctuations of scalar branes},  {\em Phys.Rev.D} {\bf 64} (2001) 064023,
  [\href{http://arxiv.org/abs/hep-th/0106041}{ hep-th/0106041}].

\bibitem{Giovannini2001b}
M.~Giovannini, {\it Thick branes and gauss-bonnet self-interactions},  {\em
  Phys.Rev.D} {\bf 64} (2001) 124004,
  [\href{http://arxiv.org/abs/hep-th/0107233}{ hep-th/0107233}].

\bibitem{Bazeia:2003aw}
D.~Bazeia, C.~Furtado, and A.~R. Gomes, {\it {Brane structure from scalar field
  in warped space-time}},  {\em JCAP} {\bf 0402} (2004) 002,
  [\href{http://arxiv.org/abs/hep-th/0308034}{ hep-th/0308034}].

\bibitem{Liu:2009ega}
Y.-X. Liu, Y.~Zhong, and K.~Yang, {\it {Scalar-Kinetic Branes}},  {\em EPL}
  {\bf 90} (2010) 51001 [\href{http://arxiv.org/abs/0907.1952}{
  arXiv:0907.1952}].

\bibitem{Bazeia:2014poa}
D.~Bazeia, L.~Losano, R.~Menezes, G.~J. Olmo, and D.~Rubiera-Garcia, {\it
  {Thick brane in $f(R)$ gravity with Palatini dynamics}},  {\em Eur.Phys.J.}
  {\bf C75} (2015) 569, [\href{http://arxiv.org/abs/1411.0897}{
  arXiv:1411.0897}].

\bibitem{Gu2014}
B.-M. Gu, B.~Guo, H.~Yu, and Y.-X. Liu, {\it {Tensor perturbations of Palatini
  $f(\mathcal{R})$-branes}},  {\em Phys.Rev.D} {\bf 92} (2015) 024011,
  [\href{http://arxiv.org/abs/1411.3241}{ arXiv:1411.3241}].

\bibitem{Meng:2003en}
X.-H. Meng and P.~Wang, {\it {Palatini formation of modified gravity with ln R
  terms}},  {\em Phys.Lett.} {\bf B584} (2004) 1--7,
  [\href{http://arxiv.org/abs/hep-th/0309062}{ hep-th/0309062}].

\bibitem{Flanagan2004}
E.~E. Flanagan, {\it {Palatini form of 1/R gravity}},  {\em Phys.Rev.Lett.}
  {\bf 92} (2004) 071101, [\href{http://arxiv.org/abs/astro-ph/0308111}{
  astro-ph/0308111}].

\bibitem{Olmo2005b}
G.~J. Olmo, {\it {Post-Newtonian constraints on f(R) cosmologies in metric and
  Palatini formalism}},  {\em Phys.Rev.D} {\bf 72} (2005) 083505,
  [\href{http://arxiv.org/abs/gr-qc/0505135}{ gr-qc/0505135}].

\bibitem{Koivisto2006}
T.~Koivisto and H.~Kurki-Suonio, {\it Cosmological perturbations in the
  palatini formulation of modified gravity},  {\em Class.Quant.Grav.} {\bf
  23} 55-69 (2006) [\href{http://arxiv.org/abs/astro-ph/0509422}{
  astro-ph/0509422}].

\bibitem{Amarzguioui2006}
M.~Amarzguioui, O.~Elgaroy, D.~Mota, and T.~Multamaki, {\it {Cosmological
  constraints on f(r) gravity theories within the palatini approach}},  {\em
  Astron.Astrophys.} {\bf 454} (2006) 707--714,
  [\href{http://arxiv.org/abs/astro-ph/0510519}{ astro-ph/0510519}].

\bibitem{Fay:2007gg}
S.~Fay, R.~Tavakol, and S.~Tsujikawa, {\it {f(R) gravity theories in Palatini
  formalism: Cosmological dynamics and observational constraints}},  {\em Phys.Rev.D} {\bf 75} (2007) 063509,
  [\href{http://arxiv.org/abs/astro-ph/0701479}{ astro-ph/0701479}].

\bibitem{Koivisto2007a}
T.~Koivisto, {\it {Viable Palatini-f(R) cosmologies with generalized dark
  matter}},  {\em Phys.Rev.D} {\bf 76} (2007) 043527,
  [\href{http://arxiv.org/abs/0706.0974}{ arXiv:0706.0974}].

\bibitem{Barausse2007}
E.~Barausse, T.~P. Sotiriou, and J.~C. Miller, {\it Curvature singularities,
  tidal forces and the viability of palatini $f(R)$ gravity},  {\em
  Class.Quant.Grav.} {\bf 25} (2008) 105008,
  [\href{http://arxiv.org/abs/0712.1141}{ arXiv:0712.1141}].

\bibitem{Barragan:2009sq}
C.~Barragan, G.~J. Olmo, and H.~Sanchis-Alepuz, {\it {Bouncing Cosmologies in
  Palatini f(R) Gravity}},  {\em Phys.Rev.D} {\bf80} (2009) 024016,
  [\href{http://arxiv.org/abs/0907.0318}{arXiv:0907.0318}].

\bibitem{Barragan2010a}
C.~Barragan and G.~J. Olmo, {\it {Isotropic and Anisotropic Bouncing
  Cosmologies in Palatini Gravity}},  {\em Phys.Rev.D} {\bf 82} (2010) 084015,
  [\href{http://arxiv.org/abs/1005.4136}{ arXiv:1005.4136}].

\bibitem{Olmo2011b}
G.~J. Olmo, {\it {Palatini Approach to Modified Gravity: f(R) Theories and
  Beyond}},  {\em Int.J.Mod.Phys.D} {\bf 20} (2011) 413--462,
  [\href{http://arxiv.org/abs/1101.3864}{ arXiv:1101.3864}].

\bibitem{Gremm2000a}
M.~Gremm, {\it Four-dimensional gravity on a thick domain wall},  {\em Phys.Lett.B} {\bf 478} (2000) 434--438,
  [\href{http://arxiv.org/abs/hep-th/9912060}{ hep-th/9912060}].

\bibitem{Gremm2000}
M.~Gremm, {\it {Thick domain walls and singular spaces}},  {\em Phys.Rev.D}
  {\bf 62} (2000) 044017, [\href{http://arxiv.org/abs/hep-th/0002040}{
  hep-th/0002040}].

\bibitem{Adam:2007ag}
C.~Adam, N.~Grandi, J.~Sanchez-Guillen, and A.~Wereszczynski, {\it {K fields,
  compactons, and thick branes}},  {\em J. Phys.} {\bf A41} (2008) 212004,
  [\href{http://arxiv.org/abs/0711.3550}{ arXiv:0711.3550}]. [Erratum: J.
  Phys.A42,159801(2009)].

\bibitem{Fu2014a}
Q.-M. Fu, L.~Zhao, K.~Yang, B.-M. Gu, and Y.-X. Liu, {\it {Stability and
  (quasi)localization of gravitational fluctuations in an Eddington-inspired
  Born-Infeld brane system}},  {\em Phys.Rev.D} {\bf 90} (2014) 104007,
  [\href{http://arxiv.org/abs/1407.6107}{ arXiv:1407.6107}].

\bibitem{Bazeia2015}
D.~Bazeia, A.~S. Lobao, and R.~Menezes, {\it Thick brane models in generalized
  theories of gravity},  {\em Phys.Lett.B} {\bf 743} (2015) 98,
  [\href{http://arxiv.org/abs/1502.04757}{ arXiv:1502.04757}].

\bibitem{Sotiriou2010}
T.~P. Sotiriou and V.~Faraoni, {\it {f(R) Theories of Gravity}},  {\em
  Rev.Mod.Phys.} {\bf 82} (2010) 451--497,
  [\href{http://arxiv.org/abs/0805.1726}{arXiv:0805.1726}].

\bibitem{Banados2010}
M.~Banados and P.~G. Ferreira, {\it {Eddington's theory of gravity and its
  progeny}},  {\em Phys.Rev.Lett.} {\bf 105} (2010) 011101,
  [\href{http://arxiv.org/abs/1006.1769}{arXiv:1006.1769}].

\bibitem{Pani2012}
P.~Pani and T.~P. Sotiriou, {\it Surface singularities in eddington-inspired
  born-infeld gravity},  {\em Phys.Rev.Lett.} {\bf 109} (2012)
  251102, [\href{http://arxiv.org/abs/1209.2972}{ arXiv:1209.2972}].

\bibitem{BeltranJimenez:2017doy}
J.~Beltran~Jimenez, L.~Heisenberg, G.~J. Olmo, and D.~Rubiera-Garcia, {\it
  {Born--Infeld inspired modifications of gravity}},  {\em Phys.Rept.} {\bf
  727} (2018) 1--129, [\href{http://arxiv.org/abs/1704.0335}{
  arXiv:1704.03351}].

\bibitem{Randall1999a}
L.~Randall and R.~Sundrum, {\it {An Alternative to compactification}},  {\em
  Phys.Rev.Lett.} {\bf 83} (1999) 4690--4693,
  [\href{http://arxiv.org/abs/hep-th/9906064}{ hep-th/9906064}].

\bibitem{Gu:2016nyo}
B.-M. Gu, Y.-P. Zhang, H.~Yu, and Y.-X. Liu, {\it {Full linear perturbations
  and localization of gravity on $f(R,T)$ brane}},  {\em Eur.Phys.J.} {\bf
  C77} (2017) 115, [\href{http://arxiv.org/abs/1606.07169}{
  arXiv:1606.07169}].

\bibitem{Dzhunushaliev2010}
V.~Dzhunushaliev, V.~Folomeev, B.~Kleihaus, and J.~Kunz, {\it {Some thick brane
  solutions in f(R)-gravity}},  {\em JHEP} {\bf 1004} (2010) 130,
  [\href{http://arxiv.org/abs/0912.2812}{ arXiv:0912.2812}].

\bibitem{Nojiri2011}
S.~Nojiri and S.~D. Odintsov, {\it {Unified cosmic history in modified gravity:
  from F(R) theory to Lorentz non-invariant models}},  {\em Phys.Rept.} {\bf
  505} (2011) 59--144, [\href{http://arxiv.org/abs/1011.0544}{
  arXiv:1011.0544}].

\bibitem{Liu:2011am}
H.~Liu, H.~Lu, and Z.-L. Wang, {\it {f(R) Gravities, Killing Spinor Equations,
  `BPS' Domain Walls and Cosmology}},  {\em JHEP} {\bf 02} (2012) 083,
  [\href{http://arxiv.org/abs/1111.6602}{ arXiv:1111.6602}].

\bibitem{Bazeia2014a}
D.~Bazeia, J.~Lobao, A.S., R.~Menezes, A.~Y. Petrov, and A.~da~Silva, {\it
  {Braneworld solutions for $F(R)$ models with non-constant curvature}},  {\em
  Phys.Lett.B} {\bf 729} (2014) 127--135,
  [\href{http://arxiv.org/abs/1311.6294}{ arXiv:1311.6294}].

\bibitem{ZhongLiu2015}
Y.~Zhong and Y.-X. Liu, {\it Pure geometric thick $f(R)$-branes: stability and
  localization of gravity}, {\em Eur.Phys.J.}  {\bf C76} (2016) 321, [\href{http://arxiv.org/abs/1507.00630}{
  arXiv:1507.00630}].

\bibitem{Bazeia:2004dh}
D.~Bazeia and A.~R. Gomes, {\it {Bloch brane}},  {\em JHEP} {\bf 05} (2004)
  012, [\href{http://arxiv.org/abs/hep-th/0403141}{ hep-th/0403141}].

\bibitem{Aybat:2010sn}
S.~M. Aybat and D.~P. George, {\it {Stability of Scalar Fields in Warped Extra
  Dimensions}},  {\em JHEP} {\bf 09} (2010) 010,
  [\href{http://arxiv.org/abs/1006.2827}{ arXiv:1006.2827}].

\bibitem{George:2011tn}
D.~P. George, {\it {Survival of scalar zero modes in warped extra dimensions}},
   {\em Phys.Rev.D} {\bf 83} (2011) 104025,
  [\href{http://arxiv.org/abs/1102.0564}{ arXiv:1102.0564}].

\bibitem{Chen:2017diy}
F.-W. Chen, B.-M. Gu, and Y.-X. Liu, {\it {Stability of braneworlds with
  non-minimally coupled multi-scalar fields}}, {\em Eur.Phys.J.} {\bf C78} (2018) 131,
  [\href{http://arxiv.org/abs/1702.03497}{ arXiv:1702.03497}].

\end{thebibliography}

\end{document}